%
%
%
%
%
%
%
%
\def\standardrisposta{s }\def\reducedrisposta{r }
\def\mplarisposta{mpla }\def\zerorisposta{z }
\def\doublerisposta{d }\def\cartarisposta{e }\def\amsrisposta{y }
\newcount\ingrandimento \newcount\sinnota \newcount\dimnota
\newcount\unoduecol \newdimen\collhsize \newdimen\tothsize
\newdimen\fullhsize \newcount\controllorisposta \sinnota=1
\newskip\infralinea  \global\controllorisposta=0
%
%
%
%
\def\risposta{s }
\def\srisposta{e }
\def\arisposta{y }
\ifx\risposta\standardrisposta \ingrandimento=1200
\message {>> This will come out UNREDUCED << }
\dimnota=2 \unoduecol=1 \global\controllorisposta=1 \fi
\ifx\risposta\reducedrisposta \ingrandimento=1095 \dimnota=1
\unoduecol=1  \global\controllorisposta=1
\message {>> This will come out REDUCED << } \fi
\ifx\risposta\doublerisposta \ingrandimento=1000 \dimnota=2
\unoduecol=2   \message {>> You must print this in
LANDSCAPE orientation << } \global\controllorisposta=1 \fi
\ifx\risposta\mplarisposta \ingrandimento=1000 \dimnota=1
\message {>> Mod. Phys. Lett. A format << }
\unoduecol=1 \global\controllorisposta=1 \fi
\ifx\risposta\zerorisposta \ingrandimento=1000 \dimnota=2
\message {>> Zero Magnification format << }
\unoduecol=1 \global\controllorisposta=1 \fi
\ifnum\controllorisposta=0  \ingrandimento=1200
\message {>>> ERROR IN INPUT, I ASSUME STANDARD
UNREDUCED FORMAT <<< }  \dimnota=2 \unoduecol=1 \fi
\magnification=\ingrandimento
%
%
%
%
\newdimen\eucolumnsize \newdimen\eudoublehsize \newdimen\eudoublevsize
\newdimen\uscolumnsize \newdimen\usdoublehsize \newdimen\usdoublevsize
\newdimen\eusinglehsize \newdimen\eusinglevsize \newdimen\ussinglehsize
\newskip\standardbaselineskip \newdimen\ussinglevsize
\newskip\reducedbaselineskip \newskip\doublebaselineskip
\eucolumnsize=12.0truecm    
\eudoublehsize=25.5truecm   
\eudoublevsize=6.7truein    
\uscolumnsize=4.4truein     
\usdoublehsize=9.4truein    
\usdoublevsize=6.8truein    
\eusinglehsize=6.5truein    
\eusinglevsize=24truecm     
\ussinglehsize=6.5truein    
\ussinglevsize=8.9truein    
\standardbaselineskip=16pt plus.2pt  
\reducedbaselineskip=14pt plus.2pt   
\doublebaselineskip=12pt plus.2pt    
%
%
\def\Portoffset{}
\def\Landoffset{\voffset=-.2truein}
\ifx\risposta\mplarisposta \def\Portoffset{\hoffset=1.8truecm} \fi
%
%
\def\Landspec{}
\tolerance=10000
\parskip=0pt plus2pt  \leftskip=0pt \rightskip=0pt
%
%
\ifx\risposta\standardrisposta \infralinea=\standardbaselineskip \fi
\ifx\risposta\reducedrisposta  \infralinea=\reducedbaselineskip \fi
\ifx\risposta\doublerisposta   \infralinea=\doublebaselineskip \fi
\ifx\risposta\mplarisposta     \infralinea=13pt \fi
\ifx\risposta\zerorisposta     \infralinea=12pt plus.2pt\fi
\ifnum\controllorisposta=0    \infralinea=\standardbaselineskip \fi
\ifx\risposta\doublerisposta   \Landoffset \else \Portoffset \fi
\ifx\risposta\doublerisposta \ifx\srisposta\cartarisposta
\tothsize=\eudoublehsize \collhsize=\eucolumnsize
\vsize=\eudoublevsize  \else  \tothsize=\usdoublehsize
\collhsize=\uscolumnsize \vsize=\usdoublevsize \fi \else
\ifx\srisposta\cartarisposta \tothsize=\eusinglehsize
\vsize=\eusinglevsize \else  \tothsize=\ussinglehsize
\vsize=\ussinglevsize \fi \collhsize=4.4truein \fi
\ifx\risposta\mplarisposta \tothsize=5.0truein
\vsize=7.8truein \collhsize=4.4truein \fi
%
%
%
%
\newcount\contaeuler \newcount\contacyrill \newcount\contaams
\font\ninerm=cmr9  \font\eightrm=cmr8  \font\sixrm=cmr6
\font\ninei=cmmi9  \font\eighti=cmmi8  \font\sixi=cmmi6
\font\ninesy=cmsy9  \font\eightsy=cmsy8  \font\sixsy=cmsy6
\font\ninebf=cmbx9  \font\eightbf=cmbx8  \font\sixbf=cmbx6
\font\ninett=cmtt9  \font\eighttt=cmtt8  \font\nineit=cmti9
\font\eightit=cmti8 \font\ninesl=cmsl9  \font\eightsl=cmsl8
\skewchar\ninei='177 \skewchar\eighti='177 \skewchar\sixi='177
\skewchar\ninesy='60 \skewchar\eightsy='60 \skewchar\sixsy='60
\hyphenchar\ninett=-1 \hyphenchar\eighttt=-1 \hyphenchar\tentt=-1
\def\bfmath{\cmmib}                 
\font\tencmmib=cmmib10  \newfam\cmmibfam  \skewchar\tencmmib='177
\font\tencmbsy=cmbsy10  \newfam\cmbsyfam  \skewchar\tencmbsy='60
\def\scaps{\cmcsc}                 
\font\tencmcsc=cmcsc10  \newfam\cmcscfam
\ifnum\ingrandimento=1095

\font\capsone=cmcsc10 at 10.95pt 

\else

\font\capsone=cmcsc10 at 12pt 
\fi

\def\ttaarr{\bf}		
\def\ppaarr{\sl}		

%
%
%
\newfam\eufmfam \newfam\msamfam \newfam\msbmfam \newfam\eufbfam
\def\Loadeulerfonts{\global\contaeuler=1 \ifx\arisposta\amsrisposta
\font\teneufm=eufm10              
\font\eighteufm=eufm8 \font\nineeufm=eufm9 \font\sixeufm=eufm6
\font\seveneufm=eufm7  \font\fiveeufm=eufm5
\font\teneufb=eufb10              
\font\eighteufb=eufb8 \font\nineeufb=eufb9 \font\sixeufb=eufb6
\font\seveneufb=eufb7  \font\fiveeufb=eufb5
\font\teneurm=eurm10              
\font\eighteurm=eurm8 \font\nineeurm=eurm9
\font\teneurb=eurb10              
\font\eighteurb=eurb8 \font\nineeurb=eurb9
\font\teneusm=eusm10              
\font\eighteusm=eusm8 \font\nineeusm=eusm9
\font\teneusb=eusb10              
\font\eighteusb=eusb8 \font\nineeusb=eusb9
\else \def\eufm{\tt} \def\eufb{\tt} \def\eurm{\tt} \def\eurb{\tt}
\def\eusm{\tt} \def\eusb{\tt}    \fi}
\def\loadeuler{\Loadeulerfonts\tenpoint}
\def\loadamsmath{\global\contaams=1 \ifx\arisposta\amsrisposta
\font\tenmsam=msam10 \font\ninemsam=msam9 \font\eightmsam=msam8
\font\sevenmsam=msam7 \font\sixmsam=msam6 \font\fivemsam=msam5
\font\tenmsbm=msbm10 \font\ninemsbm=msbm9 \font\eightmsbm=msbm8
\font\sevenmsbm=msbm7 \font\sixmsbm=msbm6 \font\fivemsbm=msbm5
\else \def\msbm{\bf} \fi \def\Bbb{\msbm} \def\symbl{\msam} \tenpoint}
\def\loadcyrill{\global\contacyrill=1 \ifx\arisposta\amsrisposta
\font\tenwncyr=wncyr10 \font\ninewncyr=wncyr9 \font\eightwncyr=wncyr8
\font\tenwncyb=wncyr10 \font\ninewncyb=wncyr9 \font\eightwncyb=wncyr8
\font\tenwncyi=wncyr10 \font\ninewncyi=wncyr9 \font\eightwncyi=wncyr8
\else \def\cyrill{\sl} \def\cyrilb{\sl} \def\cyrili{\sl} \fi\tenpoint}
\ifx\arisposta\amsrisposta
\font\sevenex=cmex7               
\font\eightex=cmex8  \font\nineex=cmex9
\font\ninecmmib=cmmib9   \font\eightcmmib=cmmib8
\font\sevencmmib=cmmib7 \font\sixcmmib=cmmib6
\font\fivecmmib=cmmib5   \skewchar\ninecmmib='177
\skewchar\eightcmmib='177  \skewchar\sevencmmib='177
\skewchar\sixcmmib='177   \skewchar\fivecmmib='177
%
%
%
\def\ninecmbsy{\tencmbsy}
\def\eightcmbsy{\tencmbsy}
\def\sevencmbsy{\tencmbsy}
\def\sixcmbsy{\tencmbsy}
\def\fivecmbsy{\tencmbsy}
\font\ninecmcsc=cmcsc9    \font\eightcmcsc=cmcsc8     \else
\def\cmmib{\fam\cmmibfam\tencmmib}\textfont\cmmibfam=\tencmmib
\scriptfont\cmmibfam=\tencmmib \scriptscriptfont\cmmibfam=\tencmmib
\def\cmbsy{\fam\cmbsyfam\tencmbsy} \textfont\cmbsyfam=\tencmbsy
\scriptfont\cmbsyfam=\tencmbsy \scriptscriptfont\cmbsyfam=\tencmbsy
\scriptfont\cmcscfam=\tencmcsc \scriptscriptfont\cmcscfam=\tencmcsc
\def\cmcsc{\fam\cmcscfam\tencmcsc} \textfont\cmcscfam=\tencmcsc \fi
\catcode`@=11
\newskip\ttglue
\gdef\tenpoint{\def\rm{\fam0\tenrm}
  \textfont0=\tenrm \scriptfont0=\sevenrm \scriptscriptfont0=\fiverm
  \textfont1=\teni \scriptfont1=\seveni \scriptscriptfont1=\fivei
  \textfont2=\tensy \scriptfont2=\sevensy \scriptscriptfont2=\fivesy
  \textfont3=\tenex \scriptfont3=\tenex \scriptscriptfont3=\tenex
  \def\mcal{\fam2 \tensy}  \def\mmit{\fam1 \teni}
  \textfont\itfam=\tenit \def\it{\fam\itfam\tenit}
  \textfont\slfam=\tensl \def\sl{\fam\slfam\tensl}
  \textfont\ttfam=\tentt \scriptfont\ttfam=\eighttt
  \scriptscriptfont\ttfam=\eighttt  \def\tt{\fam\ttfam\tentt}
  \textfont\bffam=\tenbf \scriptfont\bffam=\sevenbf
  \scriptscriptfont\bffam=\fivebf \def\bf{\fam\bffam\tenbf}
     \ifx\arisposta\amsrisposta    \ifnum\contaeuler=1
  \textfont\eufmfam=\teneufm \scriptfont\eufmfam=\seveneufm
  \scriptscriptfont\eufmfam=\fiveeufm \def\eufm{\fam\eufmfam\teneufm}
  \textfont\eufbfam=\teneufb \scriptfont\eufbfam=\seveneufb
  \scriptscriptfont\eufbfam=\fiveeufb \def\eufb{\fam\eufbfam\teneufb}
  \def\eurm{\teneurm} \def\eurb{\teneurb} \def\eusm{\teneusm}
  \def\eusb{\teneusb}    \fi    \ifnum\contaams=1
  \textfont\msamfam=\tenmsam \scriptfont\msamfam=\sevenmsam
  \scriptscriptfont\msamfam=\fivemsam \def\msam{\fam\msamfam\tenmsam}
  \textfont\msbmfam=\tenmsbm \scriptfont\msbmfam=\sevenmsbm
  \scriptscriptfont\msbmfam=\fivemsbm \def\msbm{\fam\msbmfam\tenmsbm}
     \fi      \ifnum\contacyrill=1     \def\cyrill{\tenwncyr}
  \def\cyrilb{\tenwncyb}  \def\cyrili{\tenwncyi}         \fi
  \textfont3=\tenex \scriptfont3=\sevenex \scriptscriptfont3=\sevenex
  \def\cmmib{\fam\cmmibfam\tencmmib} \scriptfont\cmmibfam=\sevencmmib
  \textfont\cmmibfam=\tencmmib  \scriptscriptfont\cmmibfam=\fivecmmib
  \def\cmbsy{\fam\cmbsyfam\tencmbsy} \scriptfont\cmbsyfam=\sevencmbsy
  \textfont\cmbsyfam=\tencmbsy  \scriptscriptfont\cmbsyfam=\fivecmbsy
  \def\cmcsc{\fam\cmcscfam\tencmcsc} \scriptfont\cmcscfam=\eightcmcsc
  \textfont\cmcscfam=\tencmcsc \scriptscriptfont\cmcscfam=\eightcmcsc
     \fi            \tt \ttglue=.5em plus.25em minus.15em
  \normalbaselineskip=12pt
  \setbox\strutbox=\hbox{\vrule height8.5pt depth3.5pt width0pt}
  \let\sc=\eightrm \let\big=\tenbig   \normalbaselines
  \baselineskip=\infralinea  \rm}
\gdef\ninepoint{\def\rm{\fam0\ninerm}
  \textfont0=\ninerm \scriptfont0=\sixrm \scriptscriptfont0=\fiverm
  \textfont1=\ninei \scriptfont1=\sixi \scriptscriptfont1=\fivei
  \textfont2=\ninesy \scriptfont2=\sixsy \scriptscriptfont2=\fivesy
  \textfont3=\tenex \scriptfont3=\tenex \scriptscriptfont3=\tenex
  \def\mcal{\fam2 \ninesy}  \def\mmit{\fam1 \ninei}
  \textfont\itfam=\nineit \def\it{\fam\itfam\nineit}
  \textfont\slfam=\ninesl \def\sl{\fam\slfam\ninesl}
  \textfont\ttfam=\ninett \scriptfont\ttfam=\eighttt
  \scriptscriptfont\ttfam=\eighttt \def\tt{\fam\ttfam\ninett}
  \textfont\bffam=\ninebf \scriptfont\bffam=\sixbf
  \scriptscriptfont\bffam=\fivebf \def\bf{\fam\bffam\ninebf}
     \ifx\arisposta\amsrisposta  \ifnum\contaeuler=1
  \textfont\eufmfam=\nineeufm \scriptfont\eufmfam=\sixeufm
  \scriptscriptfont\eufmfam=\fiveeufm \def\eufm{\fam\eufmfam\nineeufm}
  \textfont\eufbfam=\nineeufb \scriptfont\eufbfam=\sixeufb
  \scriptscriptfont\eufbfam=\fiveeufb \def\eufb{\fam\eufbfam\nineeufb}
  \def\eurm{\nineeurm} \def\eurb{\nineeurb} \def\eusm{\nineeusm}
  \def\eusb{\nineeusb}     \fi   \ifnum\contaams=1
  \textfont\msamfam=\ninemsam \scriptfont\msamfam=\sixmsam
  \scriptscriptfont\msamfam=\fivemsam \def\msam{\fam\msamfam\ninemsam}
  \textfont\msbmfam=\ninemsbm \scriptfont\msbmfam=\sixmsbm
  \scriptscriptfont\msbmfam=\fivemsbm \def\msbm{\fam\msbmfam\ninemsbm}
     \fi       \ifnum\contacyrill=1     \def\cyrill{\ninewncyr}
  \def\cyrilb{\ninewncyb}  \def\cyrili{\ninewncyi}         \fi
  \textfont3=\nineex \scriptfont3=\sevenex \scriptscriptfont3=\sevenex
  \def\cmmib{\fam\cmmibfam\ninecmmib}  \textfont\cmmibfam=\ninecmmib
  \scriptfont\cmmibfam=\sixcmmib \scriptscriptfont\cmmibfam=\fivecmmib
  \def\cmbsy{\fam\cmbsyfam\ninecmbsy}  \textfont\cmbsyfam=\ninecmbsy
  \scriptfont\cmbsyfam=\sixcmbsy \scriptscriptfont\cmbsyfam=\fivecmbsy
  \def\cmcsc{\fam\cmcscfam\ninecmcsc} \scriptfont\cmcscfam=\eightcmcsc
  \textfont\cmcscfam=\ninecmcsc \scriptscriptfont\cmcscfam=\eightcmcsc
     \fi            \tt \ttglue=.5em plus.25em minus.15em
  \normalbaselineskip=11pt
  \setbox\strutbox=\hbox{\vrule height8pt depth3pt width0pt}
  \let\sc=\sevenrm \let\big=\ninebig \normalbaselines\rm}
\gdef\eightpoint{\def\rm{\fam0\eightrm}
  \textfont0=\eightrm \scriptfont0=\sixrm \scriptscriptfont0=\fiverm
  \textfont1=\eighti \scriptfont1=\sixi \scriptscriptfont1=\fivei
  \textfont2=\eightsy \scriptfont2=\sixsy \scriptscriptfont2=\fivesy
  \textfont3=\tenex \scriptfont3=\tenex \scriptscriptfont3=\tenex
  \def\mcal{\fam2 \eightsy}  \def\mmit{\fam1 \eighti}
  \textfont\itfam=\eightit \def\it{\fam\itfam\eightit}
  \textfont\slfam=\eightsl \def\sl{\fam\slfam\eightsl}
  \textfont\ttfam=\eighttt \scriptfont\ttfam=\eighttt
  \scriptscriptfont\ttfam=\eighttt \def\tt{\fam\ttfam\eighttt}
  \textfont\bffam=\eightbf \scriptfont\bffam=\sixbf
  \scriptscriptfont\bffam=\fivebf \def\bf{\fam\bffam\eightbf}
     \ifx\arisposta\amsrisposta   \ifnum\contaeuler=1
  \textfont\eufmfam=\eighteufm \scriptfont\eufmfam=\sixeufm
  \scriptscriptfont\eufmfam=\fiveeufm \def\eufm{\fam\eufmfam\eighteufm}
  \textfont\eufbfam=\eighteufb \scriptfont\eufbfam=\sixeufb
  \scriptscriptfont\eufbfam=\fiveeufb \def\eufb{\fam\eufbfam\eighteufb}
  \def\eurm{\eighteurm} \def\eurb{\eighteurb} \def\eusm{\eighteusm}
  \def\eusb{\eighteusb}       \fi    \ifnum\contaams=1
  \textfont\msamfam=\eightmsam \scriptfont\msamfam=\sixmsam
  \scriptscriptfont\msamfam=\fivemsam \def\msam{\fam\msamfam\eightmsam}
  \textfont\msbmfam=\eightmsbm \scriptfont\msbmfam=\sixmsbm
  \scriptscriptfont\msbmfam=\fivemsbm \def\msbm{\fam\msbmfam\eightmsbm}
     \fi       \ifnum\contacyrill=1     \def\cyrill{\eightwncyr}
  \def\cyrilb{\eightwncyb}  \def\cyrili{\eightwncyi}         \fi
  \textfont3=\eightex \scriptfont3=\sevenex \scriptscriptfont3=\sevenex
  \def\cmmib{\fam\cmmibfam\eightcmmib}  \textfont\cmmibfam=\eightcmmib
  \scriptfont\cmmibfam=\sixcmmib \scriptscriptfont\cmmibfam=\fivecmmib
  \def\cmbsy{\fam\cmbsyfam\eightcmbsy}  \textfont\cmbsyfam=\eightcmbsy
  \scriptfont\cmbsyfam=\sixcmbsy \scriptscriptfont\cmbsyfam=\fivecmbsy
  \def\cmcsc{\fam\cmcscfam\eightcmcsc} \scriptfont\cmcscfam=\eightcmcsc
  \textfont\cmcscfam=\eightcmcsc \scriptscriptfont\cmcscfam=\eightcmcsc
     \fi             \tt \ttglue=.5em plus.25em minus.15em
  \normalbaselineskip=9pt
  \setbox\strutbox=\hbox{\vrule height7pt depth2pt width0pt}
  \let\sc=\sixrm \let\big=\eightbig \normalbaselines\rm }
\gdef\tenbig#1{{\hbox{$\left#1\vbox to8.5pt{}\right.\n@space$}}}
\gdef\ninebig#1{{\hbox{$\textfont0=\tenrm\textfont2=\tensy
   \left#1\vbox to7.25pt{}\right.\n@space$}}}
\gdef\eightbig#1{{\hbox{$\textfont0=\ninerm\textfont2=\ninesy
   \left#1\vbox to6.5pt{}\right.\n@space$}}}
\def\alternativefont#1#2{\ifx\arisposta\amsrisposta \relax \else
\xdef#1{#2} \fi}
\global\contaeuler=0 \global\contacyrill=0 \global\contaams=0
%
%
%
%
\newbox\fotlinebb \newbox\hedlinebb \newbox\leftcolumn
\gdef\makeheadline{\vbox to 0pt{\vskip-22.5pt
     \fullline{\vbox to8.5pt{}\the\headline}\vss}\nointerlineskip}
\gdef\makehedlinebb{\vbox to 0pt{\vskip-22.5pt
     \fullline{\vbox to8.5pt{}\copy\hedlinebb\hfil
     \line{\hfill\the\headline\hfill}}\vss} \nointerlineskip}
\gdef\makefootline{\baselineskip=24pt \fullline{\the\footline}}
\gdef\makefotlinebb{\baselineskip=24pt
    \fullline{\copy\fotlinebb\hfil\line{\hfill\the\footline\hfill}}}
\gdef\doubleformat{\shipout\vbox{\Landspec\makehedlinebb
     \fullline{\box\leftcolumn\hfil\columnbox}\makefotlinebb}
     \advancepageno}
\gdef\columnbox{\leftline{\pagebody}}
\gdef\line#1{\hbox to\hsize{\hskip\leftskip#1\hskip\rightskip}}
\gdef\fullline#1{\hbox to\fullhsize{\hskip\leftskip{#1}%
\hskip\rightskip}}
\gdef\footnote#1{\let\@sf=\empty
         \ifhmode\edef\#sf{\spacefactor=\the\spacefactor}\/\fi
         #1\@sf\vfootnote{#1}}
\gdef\vfootnote#1{\insert\footins\bgroup
         \ifnum\dimnota=1  \eightpoint\fi
         \ifnum\dimnota=2  \ninepoint\fi
         \ifnum\dimnota=0  \tenpoint\fi
         \interlinepenalty=\interfootnotelinepenalty
         \splittopskip=\ht\strutbox
         \splitmaxdepth=\dp\strutbox \floatingpenalty=20000
         \leftskip=\oldssposta \rightskip=\olddsposta
         \spaceskip=0pt \xspaceskip=0pt
         \ifnum\sinnota=0   \textindent{#1}\fi
         \ifnum\sinnota=1   \item{#1}\fi
         \footstrut\futurelet\next\fo@t}
\gdef\fo@t{\ifcat\bgroup\noexpand\next \let\next\f@@t
             \else\let\next\f@t\fi \next}
\gdef\f@@t{\bgroup\aftergroup\@foot\let\next}
\gdef\f@t#1{#1\@foot} \gdef\@foot{\strut\egroup}
\gdef\footstrut{\vbox to\splittopskip{}}
\skip\footins=\bigskipamount
\count\footins=1000  \dimen\footins=8in
\catcode`@=12
\tenpoint
\ifnum\unoduecol=1 \hsize=\tothsize   \fullhsize=\tothsize \fi
\ifnum\unoduecol=2 \hsize=\collhsize  \fullhsize=\tothsize \fi
\global\let\lrcol=L      \ifnum\unoduecol=1
\output{\plainoutput{\ifnum\tipbnota=2 \clearnmbnota\fi}} \fi
\ifnum\unoduecol=2 \output{\if L\lrcol
     \global\setbox\leftcolumn=\columnbox
     \global\setbox\fotlinebb=\line{\hfill\the\footline\hfill}
     \global\setbox\hedlinebb=\line{\hfill\the\headline\hfill}
     \advancepageno  \global\let\lrcol=R
     \else  \doubleformat \global\let\lrcol=L \fi
     \ifnum\outputpenalty>-20000 \else\dosupereject\fi
     \ifnum\tipbnota=2\clearnmbnota\fi }\fi
\def\ifdoublepage{\ifnum\unoduecol=2 }
\gdef\yespagenumbers{\footline={\hss\tenrm\folio\hss}}
\gdef\ciao{ \ifnum\fdefcontre=1 \endfdef\fi
     \par\vfill\supereject \ifnum\unoduecol=2
     \if R\lrcol  \headline={}\nopagenumbers\null\vfill\eject
     \fi\fi \end}

\newskip\olddsposta \newskip\oldssposta
\global\oldssposta=\leftskip \global\olddsposta=\rightskip

\def\filldots{\leaders\hbox to 1em{\hss.\hss}\hfill}
\def\inquadrb#1 {\vbox {\hrule  \hbox{\vrule \vbox {\vskip .2cm
    \hbox {\ #1\ } \vskip .2cm } \vrule  }  \hrule} }
 \def\newline{\hfil\break}
\def\jump{\vskip\baselineskip} \newskip\iinnffrr
\def\sjump{\iinnffrr=\baselineskip
          \divide\iinnffrr by 2 \vskip\iinnffrr}
\def\bjump{\vskip\baselineskip \vskip\baselineskip}
\newcount\nmbnota  \def\clearnmbnota{\global\nmbnota=0}
\newcount\tipbnota \def\letterfootnote{\global\tipbnota=1}

\def\note#1{\global\advance\nmbnota by 1 \ifnum\tipbnota=1
    \footnote{$^{\rm\nttlett}$}{#1} \else {\ifnum\tipbnota=2
    \footnote{$^{\nttsymb}$}{#1}
    \else\footnote{$^{\the\nmbnota}$}{#1}\fi}\fi}
\def\nttlett{\ifcase\nmbnota \or a\or b\or c\or d\or e\or f\or
g\or h\or i\or j\or k\or l\or m\or n\or o\or p\or q\or r\or
s\or t\or u\or v\or w\or y\or x\or z\fi}
\def\nttsymb{\ifcase\nmbnota \or\dag\or\sharp\or\ddag\or\star\or
\natural\or\flat\or\clubsuit\or\diamondsuit\or\heartsuit
\or\spadesuit\fi}   \clearnmbnota
\def\numberfootnote{\global\tipbnota=0} \numberfootnote
\def\setnote#1{\expandafter\xdef\csname#1\endcsname{
\ifnum\tipbnota=1 {\rm\nttlett} \else {\ifnum\tipbnota=2
{\nttsymb} \else \the\nmbnota\fi}\fi} }
\newcount\nbmfig  \def\clearnbmfig{\global\nbmfig=0}
\gdef\figure{\global\advance\nbmfig by 1
      {\rm fig. \the\nbmfig}}   \clearnbmfig
\def\setfig#1{\expandafter\xdef\csname#1\endcsname{fig. \the\nbmfig}}
 \def\endformula{\eqno\numero $$}
 \def\efr{\endformula}
\newcount\frmcount \def\clearfrmcount{\global\frmcount=0}
\def\numero{\global\advance\frmcount by 1   \ifnum\indappcount=0
  {\ifnum\cpcount <1 {\hbox{\rm (\the\frmcount )}}  \else
  {\hbox{\rm (\the\cpcount .\the\frmcount )}} \fi}  \else
  {\hbox{\rm (\applett .\the\frmcount )}} \fi}
\def\nameformula#1{\global\advance\frmcount by 1%
\ifnum\draftnum=0  {\ifnum\indappcount=0%
{\ifnum\cpcount<1\xdef\spzzttrra{(\the\frmcount )}%
\else\xdef\spzzttrra{(\the\cpcount .\the\frmcount )}\fi}%
\else\xdef\spzzttrra{(\applett .\the\frmcount )}\fi}%
\else\xdef\spzzttrra{(#1)}\fi%
\expandafter\xdef\csname#1\endcsname{\spzzttrra}
\eqno \hbox{\rm\spzzttrra} $$}
\def\nfr{\nameformula}    
\def\nameali#1{\global\advance\frmcount by 1%
\ifnum\draftnum=0  {\ifnum\indappcount=0%
{\ifnum\cpcount<1\xdef\spzzttrra{(\the\frmcount )}%
\else\xdef\spzzttrra{(\the\cpcount .\the\frmcount )}\fi}%
\else\xdef\spzzttrra{(\applett .\the\frmcount )}\fi}%
\else\xdef\spzzttrra{(#1)}\fi%
\expandafter\xdef\csname#1\endcsname{\spzzttrra}
  \hbox{\rm\spzzttrra} }      \clearfrmcount
\newcount\cpcount \def\clearcpcount{\global\cpcount=0}
\newcount\subcpcount \def\clearsubcpcount{\global\subcpcount=0}
\newcount\appcount \def\clearappcount{\global\appcount=0}
\newcount\indappcount \def\clearindappcount{\indappcount=0}
\newcount\sottoparcount 

\def\applett{\ifcase\appcount  \or {A}\or {B}\or {C}\or
{D}\or {E}\or {F}\or {G}\or {H}\or {I}\or {J}\or {K}\or {L}\or
{M}\or {N}\or {O}\or {P}\or {Q}\or {R}\or {S}\or {T}\or {U}\or
{V}\or {W}\or {X}\or {Y}\or {Z}\fi    \ifnum\appcount<0
\immediate\write16 {Panda ERROR - Appendix: counter "appcount"
out of range}\fi  \ifnum\appcount>26  \immediate\write16 {Panda
ERROR - Appendix: counter "appcount" out of range}\fi}
\clearappcount  \clearindappcount \newcount\connttrre
\def\clearconnttrre{\global\connttrre=0} \newcount\countref
\def\clearcountref{\global\countref=0} \clearcountref
\def\chapter#1{\global\advance\cpcount by 1 \clearfrmcount
                 \goodbreak\null\vbox{\jump\nobreak
                 \clearsubcpcount\clearindappcount
                 \itemitem{\ttaarr\the\cpcount .\qquad}{\ttaarr #1}
                 \par\nobreak\jump\sjump}\nobreak}
\def\section#1{\global\advance\subcpcount by 1 \goodbreak\null
               \vbox{\sjump\nobreak\ifnum\indappcount=0
                 {\ifnum\cpcount=0 {\itemitem{\ppaarr
               .\the\subcpcount\quad\enskip\ }{\ppaarr #1}\par} \else
                 {\itemitem{\ppaarr\the\cpcount .\the\subcpcount\quad
                  \enskip\ }{\ppaarr #1} \par}  \fi}
                \else{\itemitem{\ppaarr\applett .\the\subcpcount\quad
                 \enskip\ }{\ppaarr #1}\par}\fi\nobreak\jump}\nobreak}
\clearsubcpcount
\def\appendix#1{\global\advance\appcount by 1 \clearfrmcount
                  \goodbreak\null\vbox{\jump\nobreak
                  \global\advance\indappcount by 1 \clearsubcpcount
          \itemitem{ }{\hskip-40pt\ttaarr Appendix\ #1}
             \nobreak\jump\sjump}\nobreak}
\clearappcount \clearindappcount
\def\references{\goodbreak\null\vbox{\jump\nobreak
   \itemitem{}{\ttaarr References} \nobreak\jump\sjump}\nobreak}

\clearcpcount\clearcountref

\def\setchap#1{\ifnum\indappcount=0{\ifnum\subcpcount=0%
\xdef\spzzttrra{\the\cpcount}%
\else\xdef\spzzttrra{\the\cpcount .\the\subcpcount}\fi}
\else{\ifnum\subcpcount=0 \xdef\spzzttrra{\applett}%
\else\xdef\spzzttrra{\applett .\the\subcpcount}\fi}\fi
\expandafter\xdef\csname#1\endcsname{\spzzttrra}}
\newcount\draftnum \newcount\ppora   \newcount\ppminuti
\global\ppora=\time   \global\ppminuti=\time
\global\divide\ppora by 60  \draftnum=\ppora
\multiply\draftnum by 60    \global\advance\ppminuti by -\draftnum
\def\droggi{\number\day /\number\month /\number\year\ \the\ppora
:\the\ppminuti}     \global\draftnum=0
\def\draftcomment#1{\ifnum\draftnum=0 \relax \else
{\ {\bf ***}\ #1\ {\bf ***}\ }\fi} 
%
%
\catcode`@=11
\gdef\Ref#1{\expandafter\ifx\csname @rrxx@#1\endcsname\relax%
{\global\advance\countref by 1    \ifnum\countref>200
\immediate\write16 {Panda ERROR - Ref: maximum number of references
exceeded}  \expandafter\xdef\csname @rrxx@#1\endcsname{0}\else
\expandafter\xdef\csname @rrxx@#1\endcsname{\the\countref}\fi}\fi
\ifnum\draftnum=0 \csname @rrxx@#1\endcsname \else#1\fi}
\gdef\beginref{\ifnum\draftnum=0  \gdef\Rref{\fairef}
\gdef\endref{\scriviref} \else\relax\fi
\ifx\risposta\mplarisposta \ninepoint \fi
\parskip 2pt plus.2pt \baselineskip=12pt}
\def\Reflab#1{[#1]} \gdef\Rref#1#2{\item{\Reflab{#1}}{#2}}
\gdef\endref{\relax}  \newcount\conttemp
\gdef\fairef#1#2{\expandafter\ifx\csname @rrxx@#1\endcsname\relax
{\global\conttemp=0 \immediate\write16 {Panda ERROR - Ref: reference
[#1] undefined}} \else
{\global\conttemp=\csname @rrxx@#1\endcsname } \fi
\global\advance\conttemp by 50  \global\setbox\conttemp=\hbox{#2} }
\gdef\scriviref{\clearconnttrre\conttemp=50
\loop\ifnum\connttrre<\countref \advance\conttemp by 1
\advance\connttrre by 1
\item{\Reflab{\the\connttrre}}{\unhcopy\conttemp} \repeat}
\clearcountref \clearconnttrre
\catcode`@=12
\ifx\risposta\mplarisposta \def\Reflab#1{#1.} \letterfootnote \fi

\def\slashchar#1{\setbox0=\hbox{$#1$} \dimen0=\wd0
     \setbox1=\hbox{/} \dimen1=\wd1 \ifdim\dimen0>\dimen1
      \rlap{\hbox to \dimen0{\hfil/\hfil}} #1 \else
      \rlap{\hbox to \dimen1{\hfil$#1$\hfil}} / \fi}
\ifx\oldchi\undefined \let\oldchi=\chi
  \def\cchi{{\raise 1pt\hbox{$\oldchi$}}} \let\chi=\cchi \fi

\def\frac#1#2{{\textstyle{#1 \over #2}}}

\def\half{\ifinner {\scriptstyle {1 \over 2}}\else {1 \over 2} \fi}

\def\simge{\rlap{\raise 2pt \hbox{$>$}}{\lower 2pt \hbox{$\sim$}}}
\def\simle{\rlap{\raise 2pt \hbox{$<$}}{\lower 2pt \hbox{$\sim$}}}

\def\vbig#1#2{{\vbigd@men=#2\divide\vbigd@men by 2%
\hbox{$\left#1\vbox to \vbigd@men{}\right.\n@space$}}}

%
%
\newcount\fdefcontre \newcount\fdefcount \newcount\indcount
\newread\filefdef  \newread\fileftmp  \newwrite\filefdef
\newwrite\fileftmp     \def\strip#1*.A {#1}
\def\futuredef#1{\beginfdef
\expandafter\ifx\csname#1\endcsname\relax%
{\immediate\write\fileftmp {#1*.A}
\immediate\write16 {Panda Warning - fdef: macro "#1" on page
\the\pageno \space undefined}
\ifnum\draftnum=0 \expandafter\xdef\csname#1\endcsname{(?)}
\else \expandafter\xdef\csname#1\endcsname{(#1)} \fi
\global\advance\fdefcount by 1}\fi   \csname#1\endcsname}

\def\beginfdef{\ifnum\fdefcontre=0
\immediate\openin\filefdef \jobname.fdef
\immediate\openout\fileftmp \jobname.ftmp
\global\fdefcontre=1  \ifeof\filefdef \immediate\write16 {Panda
WARNING - fdef: file \jobname.fdef not found, run TeX again}
\else \immediate\read\filefdef to\spzzttrra
\global\advance\fdefcount by \spzzttrra
\indcount=0      \loop\ifnum\indcount<\fdefcount
\advance\indcount by 1   \immediate\read\filefdef to\spezttrra
\immediate\read\filefdef to\sppzttrra
\edef\spzzttrra{\expandafter\strip\spezttrra}
\immediate\write\fileftmp {\spzzttrra *.A}
\expandafter\xdef\csname\spzzttrra\endcsname{\sppzttrra}
\repeat \fi \immediate\closein\filefdef \fi}
\def\endfdef{\immediate\closeout\fileftmp   \ifnum\fdefcount>0
\immediate\openin\fileftmp \jobname.ftmp
\immediate\openout\filefdef \jobname.fdef
\immediate\write\filefdef {\the\fdefcount}   \indcount=0
\loop\ifnum\indcount<\fdefcount    \advance\indcount by 1
\immediate\read\fileftmp to\spezttrra
\edef\spzzttrra{\expandafter\strip\spezttrra}
\immediate\write\filefdef{\spzzttrra *.A}
\edef\spezttrra{\string{\csname\spzzttrra\endcsname\string}}
\iwritel\filefdef{\spezttrra}
\repeat  \immediate\closein\fileftmp \immediate\closeout\filefdef
\immediate\write16 {Panda Warning - fdef: Label(s) may have changed,
re-run TeX to get them right}\fi}
\def\iwritel#1#2{\newlinechar=-1
{\newlinechar=`\ \immediate\write#1{#2}}\newlinechar=-1}
\global\fdefcontre=0 \global\fdefcount=0 \global\indcount=0
%
%
\null
%
%
%
%


%
\loadamsmath
\loadeuler
\mathchardef\bbalpha="710B
\mathchardef\bbbeta="710C
\mathchardef\bbgamma="710D
\mathchardef\bbdelta="710E
\mathchardef\bbxi="7118
\mathchardef\bbpsi="7112
\mathchardef\bbomega="7121
\mathchardef\sdir="2D6E
\mathchardef\dirs="2D6F
\def\bal{{\bfmath\bbalpha}}
\def\bb{{\bfmath\bbbeta}}
\def\bgamma{{\bfmath\bbgamma}}
\def\bd{{\bfmath\bbdelta}}

\def\ba{{\bfmath a}}

\def\bH{{\bfmath H}}

\def\ur{{\underline r}}

\def\thalf{{\scriptstyle {3 \over 2}}}
\pageno=0\baselineskip=14pt
\nopagenumbers{
\line{\hfill SWAT/135}
\line{\hfill\tt hep-th/9611106}
\line{\hfill November 1996}
\ifdoublepage \bjump\bjump\bjump\bjump\else\vfill\fi
\centerline{\capsone Semi-classical decay of monopoles in $N=2$ gauge theory}
\bjump\sjump
\centerline{\scaps  Timothy J. Hollowood}
\sjump
\centerline{\sl Department of Physics, University of Wales Swansea,}
\centerline{\sl Singleton Park, Swansea SA2 8PP, U.K.}
\centerline{\tt  t.hollowood@swansea.ac.uk}
\sjump
\bjump\bjump
\ifdoublepage
\vfill
\eject\null\vfill\fi
\centerline{\capsone ABSTRACT}\sjump
It is shown how monopoles and dyons decay on curves of marginal
stability in the moduli space of vacua at weak coupling in pure
$N=2$ gauge theory with arbitrary gauge group. The analysis involves
a semi-classical treatment of the monopole and rests on the fact that
the monopole moduli space spaces for a magnetic charge vector equal to a 
non-simple root
enlarge discontinuously at the curves of marginal stability. This
enlargement of the moduli space describes the freedom for the monopole
to be separated into stable constituent monopoles. Such decays do not
occur in the associated theory with $N=4$ supersymmetry because in
this case there exist bound-states at threshold.
\sjump\vfill
\eject}
 \vfill

\yespagenumbers\pageno=1
%
%

\chapter{Introduction}

The recent progress in
understanding supersymmetric gauge theories with more than one
supersymmetry, promises to lead to a complete
understanding of the spectrum of BPS-saturated
states (or BPS states, for short). What is particularly fascinating is
that the spectrum of these states can also be probed by
semi-classical methods, thus yielding a
way to test the validity of the exact solutions proposed. 

One of the most novel phenomenon in these theories 
is the possibility for the decay
of BPS states in certain regions of the moduli space of vacua
${\cal M}_{\rm vac}$ [\Ref{SW1},\Ref{SW2}]. Generically, 
these decays occur on curves of co-dimension
one (for $N=2$ supersymmetry) 
and so ${\cal M}_{\rm vac}$ is divided
into regions separated by the decay curves. Since these decays always occur
for kinematical reasons at threshold the decay curves are known as
Curves of Marginal Stability (CMS). In principle, the spectrum of BPS
states in the theory can be different in each region, with the
discrepancies being due to decay processes across the CMS.

In pure $N=2$ supersymmetric SU(2) gauge theory there is a CMS which
occurs only in the strong coupling region
[\Ref{SW1},\Ref{F1},\Ref{AFS}]. Bilal and
Ferrari [\Ref{BF}] have shown using an
ingenious symmetry argument that certain dyons do actually decay on the CMS. 
When matter is added in the form of hypermultiplets,
it has been pointed out by Seiberg and Witten [\Ref{SW2}] that a CMS will occur
under certain conditions (when the mass of a hypermultiplet
and the Higgs VEV are much greater than $\Lambda$)
even in the region of weak coupling and hence
should be describable within a semi-classical approach. 
In this paper we will study another example where CMS extend into the
region of weak coupling, namely 
pure $N=2$ gauge theory with
larger gauge groups [\Ref{FH1}]. 
In these theories we answer the question as to how 
monopole and dyon decay is
described within the semi-classical approximation. These decays are of
the form
$$
{\rm dyon}\rightarrow{\rm dyon}+{\rm dyon}.
\efr
The case with
hypermultiplets is somewhat different and involves decays of the form
$$
{\rm dyon}\rightarrow{\rm dyon}+{\rm quark},
\efr
and has been investigated in [\Ref{H1}].

Of central important in the  semi-classical formalism for quantizing
monopoles is the
moduli space of the monopole solution ${\cal M}_{\rm
mon}$ (not to be confused with the moduli space of vacua ${\cal
M}_{\rm vac}$). In fact,
the calculation of the semi-classical spectrum of dyons can be reduced to 
the problem of quantum mechanics on ${\cal M}_{\rm mon}$ [\Ref{MAN}]. 
Although, in theories with arbitrary gauge groups, the structure of 
${\cal M}_{\rm mon}$ is rather complicated, quite a lot of information
on the nature of these spaces is known. First of all, in theories with
a real adjoint Higgs field in the Prasad-Sommerfeld limit, one can construct
monopole solutions by embeddings of the 
SU(2) spherically symmetric 
't Hooft-Polyakov monopole [\Ref{TP}]. 
These solutions are associated to particular
roots of the Lie algebra of the gauge group. One can 
then investigate the form of the moduli space of these solutions
locally by an analysis of zero modes. For
technical reasons it is actually more convenient to consider the zero modes of
the Dirac equation in the back-ground of the monopole solution
[\Ref{WB1}]. The
bosonic, or Yang-Mills, zero modes are then related to the fermionic
zero modes by supersymmetry [\Ref{G1}]. 

It turns out that ${\cal M}_{\rm mon}$ always has a factor of the form
${\Bbb R}^3$. This simply reflects the freedom to move the
centre-of-mass of the monopole solution.
If the monopole is associated to a simple root\note{We shall denote
the simple roots of the Lie algebra $g$ of the gauge group as
$\bal_i$, $i=1,2,\ldots,r$.} (with respect to a
dominant Weyl chamber defined by the Higgs VEV)
then there is one additional zero mode reflecting the freedom to choose
the periodic 
U(1) ``charge angle'' of the monopole. In this case the monopole is
``fundamental'' and has a moduli space of the form
$$
{\cal M}_{\rm mon}={\Bbb R}\times S^1.
\efr
On the contrary, if the monopole is associated to a
non-simple root $\bal$, i.e. has a magnetic charge vector of the form
$$
\bal^\star=\sum_{i=1}^rn_i\bal_i^\star,
\nfr{COMP}
where $\bb^\star=\bb/\bb^2$,
then there are additional zero modes that reflect the
fact that the solution can be deformed away from spherical symmetry
[\Ref{WB1}].
In this picture such a monopole is composite and consists of
$\left(\sum_{i=1}^rn_i\right)$ fundamental monopoles. 
(We are assuming that the gauge group is broken to its maximal
abelian subgroup by the adjoint Higgs mechanism.) Asymptotic solutions 
to the equations of motion (the
Bogomol'nyi equation) can be constructed by simply
superimposing well-separated fundamental monopole solutions.
For the case when the monopole consists of a pair of distinct fundamental
monopoles, i.e. $\bal^\star=\bal_a^\star+\bal_b^\star$, for $a\neq b$,
the moduli space has the form [\Ref{LWY2}]
$$
{\cal M}_{\rm mon}={\Bbb R}^3\times{{\Bbb R}\times{\cal M}_0\over{\Bbb
Z}},
\efr
where ${\Bbb Z}$ is a normal subgroup of the isometry group of
${\Bbb R}\times{\cal M}_0$. In the above the factor of ${\Bbb R}$ is
associated to a particular linear combination of the $r={\rm rank}(g)$
unbroken U(1) gauge degrees-of-freedom and
${\cal M}_0$ is a 4 dimensional Euclidean Taub-NUT manifold. This
space was first studied for the case of the monopole associated to the
non-simple root of SU(3) in unpublished work by Connell
[\Ref{SC}]. This example was subsequently discussed by Gauntlett and
Lowe [\Ref{GL}] and Lee, Weinberg and Yi [\Ref{LWY1}], who also showed
that ${\cal M}_0$ admitted a unique middle-dimensional 
square-integrable harmonic form, thus
providing evidence for exact duality in the associated $N=4$
theory. Lee, Weinberg and Yi [\Ref{LWY2}] then went on to conjecture a
form for ${\cal M}_0$ for solutions consisting of 
distinct fundmental monopoles in any Lie algebra, 
i.e. $n_i$ equal to 0 or 1 only (this implies that the 
vector $\bal$ defined in \COMP\ must be a root).
The conjecture was proven in the case of SU$(n)$ gauge group by 
Chalmers [\Ref{GC}] and Murray [\Ref{MUR}].
The unique middle-dimensional 
square-integrable harmonic forms on these spaces were constructed in 
a general way by Gibbons [\Ref{GIB1}]. Some related work is contained in
[\Ref{LWY3},\Ref{GIB2}]. At the moment, the structure of
${\cal M}_0$ in the general case, when $\bal$ is any root and so
the $n_i$'s can be more than 1,
is not known. However, as we shall explain below, if we take as a
hypothesis that the spectrum of
$N=4$ theories exhibits exact electro-magnetic duality (or
Goddard-Nuyts-Olive (GNO) duality [\Ref{GNO}]), this implies that
for any multi-monopole solution for which the vector $\bal$, defined 
in \COMP, is a root of the Lie algebra then the
associated manifold ${\cal M}_0$ admits a unique square-integrable
harmonic form.

Both $N=2$ and $N=4$ supersymmetric gauge theories are of the type
discussed above, with adjoint Higgs fields. However, in these theories
the Higgs fields are also vectors under an SO($N_{\cal R}$) R-symmetry,
where $N_{\cal R}$ equals 2 and 6, for $N=2$ and $N=4$
supersymmetries, respectively. This will have important implications for the
form  of the monopole moduli spaces. In particular, these moduli
spaces are {\it not\/} in general identical to those theories with real
Higgs because the additional 
zero modes corresponding to the factor ${\cal M}_0$ are
generically stablized. However, on the CMS in ${\cal M}_{\rm vac}$
some of the additonal zero modes survive. 
So ${\cal
M}_0$, which is generically trivial, can become a non-trivial
submanifold of the moduli space of the same monopole in the real Higgs
theory.

The mechanism for monopole decay is now revealed: generically a
monopole (or dyon) associated to any root of the Lie
algebra is stable; however, on a CMS its moduli space enlarges
discontinuously and the monopole may decay into stable
constituents. The condition that the monopole remains stable requires
the existence of a bound-state at threshold which is described in the
semi-classical formalism by the existence of a certain
differential form on ${\cal M}_0$. For an $N=4$ theory the form must
be square-integrable and harmonic, 
whilst in an $N=2$ theory the form must be a square-integrable purely
holomorphic harmonic form. (It turns out that ${\cal M}_0$ is
always a hyper-K\"ahler manifold, see for example [\Ref{G1}].)
This means that a knowledge of the spectrum
of the $N=4$ theory has implications for the decay of dyons in the
$N=2$ theory. More specifically if we make the assumption that
GNO duality of the spectrum of BPS states 
is exact in the $N=4$ theory
then this implies that ${\cal M}_0$ always admits a unique
square-integrable harmonic
form. The uniqueness means that the form cannot be holomorphic,
since otherwise there would be an anti-holomorphic partner. So exact
GNO duality in the $N=4$ theory implies that dyons associated to
non-simple roots {\it actually\/} decay on their CMS in the $N=2$
theory (at least in pure gauge theories with no matter).
Of course, the weakness in this chain of argument is the assumption
that the spectrum of the $N=4$ theory has exact GNO duality following
from a lack of knowledge of the structure of ${\cal
M}_0$. However, the results that are known in the literature are
enough to prove the result for all the simply-laced Lie groups. In the
other cases, we will take the assumption of exact duality in $N=4$
theories to be a working hypothesis.

\chapter{Embedding monopole solutions in vector Higgs models}

In this section we show how to embed the SU(2) 't Hooft-Polyakov
monopole solution in a theory which has $N=2$ or $N=4$ supersymmetry
and an arbitrary gauge group. The case of $N=2$, where the Higgs field
is complex has been discussed in [\Ref{AY}]. In these theories the
Higgs field $\Phi^I$ is adjoint valued and also carries an additional
vector index $I=1,2,\ldots,N_{\cal R}$, where $N_{\cal R}$ equals 2
and 6, for $N=2$ and $N=4$ supersymmetry, respectively.    
The fermion fields will play no role in the construction
of the monopole solutions and so we will ignore them for the rest of
this section.

The first task is to 
establish the form of the Bogomol'nyi bound in a vector
Higgs model. The energy of a configuration involving the bosonic fields is 
$$ 
U={1\over2g}\int d^3x\,{\rm Tr}\left(E_i^2+B_i^2+
\left(D_0\Phi^I\right)^2+\left(D_i\Phi^I\right)^2
+\sum_{I<J}\left[\Phi^I,\Phi^J\right]^2\right).
\nfr{ENER}
Although we have ignored the fermion fields, supersymmetry leaves its
mark in the very special form of the Higgs potential in
\ENER. The energy can be re-expressed in the following form
$$ 
U={1\over2g}\int d^3x\,{\rm Tr}\left(E_i^2+\left(\lambda^IB_i-D_i
\Phi^I\right)^2+\left(D_0\Phi^I\right)^2+2B_i\lambda^ID_i\Phi^I+
\sum_{I<J}\left[\Phi^I,\Phi^J\right]^2\right).
\nfr{ENT}
where $\lambda^I$ is at this stage 
an arbitrary constant vector of unit length.\note{We will
denote the length of an SO($N_{\cal R}$) vector as $\vert\lambda^I\vert$.}

From \ENT\ we derive a bound for the energy of a configuration
$$
U\geq{4\pi\over g}\lambda^IQ_M^I,
\efr
where the magnetic charge is defined to be
$$
Q_M^I={1\over 4\pi}\int dS_i\,{\rm Tr}\left(B_i\Phi^I\right),
\efr
with the integral defined over a large sphere at spatial infinity.
The most stringent bound is achieved by choosing $\lambda^I=
Q_M^I/\vert Q_M^I\vert$, giving
$$
U\geq{4\pi\over g}\left\vert Q_M^I\right\vert.
\efr
This is the analogue of the famous Bogomol'nyi bound for magnetically
charged configurations in this model. A configuration which saturates the
bound must satisfy the equations
$$
D_0\Phi^I=0,\qquad E_i=0,\qquad \lambda^IB_i=D_i\Phi^I,\qquad[\Phi^I,\Phi^J]=0.
\nfr{BOGE}
where
$$
\lambda^I={Q_M^I\over\left\vert Q_M^I\right\vert}.
\nfr{DEFL}

We now show how the SU(2) 't Hooft-Polyakov monopole [\Ref{TP}], in the
Prasad-Sommerfeld limit [\Ref{PS}], may be embedded
in our model to produce solutions to \BOGE.
For convenience we shall work in a unitary gauge
where the VEV of the Higgs field on the sphere at spatial 
infinity is a constant adjoint-valued SO($N_{\cal R}$) vector $a^I$. 
Since for the VEV $[a^I,a^J]=0$, the components $a^I$ can be simultaneously
diagonalized by a global gauge transformation. Denoting the Cartan generators
of $g$ by the $r={\rm rank}(g)$ vector $\bH$ we have
$$
\lim_{\vert{\underline r}
\vert\rightarrow\infty}\Phi^I({\underline r})=a^I=\ba^I\cdot\bH,
\efr
which defines the $N_{\cal R}$ $r$-dimensional vectors
$\ba^I$. However, this does not completely fix the gauge symmetry
since it leaves the freedom to perform discrete gauge transformations
in the Weyl group of $G$. This freedom can be fixed, for example,
by demanding that
$\ba^1$ is in the fundamental Weyl chamber with respect to some
choice of simple roots $\bal_i$:
$$
\ba^1\cdot\bal_i\geq0,\qquad i=1,2,\ldots,r.
\efr
This defines the classical moduli space of vacua $W$. For theories
with $N=2$ supersymmetry, the quantum
moduli space of vacua will be approximately $W$ in the
region of weak coupling, that is at distances much greater than
$\Lambda$, the scale of strong coupling effects, away from the subspaces
where $\ba^I\cdot\bal_i=0$. These are precisely the subspaces where
classically one would expect enhanced gauge symmetries. In the $N=4$ theories,
we would not expect strong coupling effects to appear and the quantum
moduli space should equal $W$ for any value of the coupling. In any
case, in what follows 
we will explicitly work away from the regions $\ba^I\cdot\bal_i=0$ and
hence avoid some of the problems associated with monopoles in theories
with long-range non-abelian fields [\Ref{DFHK1}].

Let the fields of the SU(2) BPS monopole solution be 
$$
\phi(\ur)=\phi_j(\ur;\lambda)t_j,\qquad
A(\ur)=A_j(\ur;\lambda)t_j,\qquad A_0=0.
\efr
In unitary gauge the Higgs VEV on the sphere at spatial infinity is (say)
$$
\lim_{\vert{\underline r}\vert\rightarrow\infty}\phi(\ur)=\lambda t_3.
\efr
In the above, $t_j$ are generators of SU(2) normalized so that 
$[t_i,t_j]=i\epsilon_{ijk}t_k$. This solution satisfies the
Bogomol'nyi equation
$$
B_i=D_i\phi.
\efr

We now wish to embed this solution in
the vector Higgs model with gauge group $G$. The procedure is a
generalization of the embeddings of SU(2) monopoles in higher gauge
groups with a single real Higgs field discussed by Weinberg [\Ref{WB1}]
following the earlier work of Bais [\Ref{BAIS}].
Take an embedding of $su(2)$ in the Lie algebra $g$ given by
$$\eqalign{
t_1&=(2\bal^2)^{-1/2}\left(E_\bal+E_{-\bal}\right)\cr
t_2&=-i(2\bal^2)^{-1/2}\left(E_\bal-E_{-\bal}\right)\cr
t_3&=\bal\cdot\bH/\bal^2,\cr}
\nfr{SUT}
where $\bal$ is some root of $g$. In \SUT\ we have used a Cartan-Weyl
basis for $g$.

The embedded spherically symmetric monopole solution is given by
$$
\Phi^I(\ur)=\lambda^I\phi_j\left(\ur;\vert\ba^I\cdot\bal\vert\right)t_j+\xi^I,
\qquad A(\ur)=A_j(\ur;\lambda)t_j,
\efr
where $\lambda^I$ is defined in \DEFL\ and so equals
$$
\lambda^I={\ba^I\cdot\bal\over\left\vert\ba^I\cdot\bal\right\vert},
\efr
and 
$\xi^I$ is the constant adjoint-valued vector
$$
\xi^I=\left(\ba^I-{\ba^I\cdot\bal\over\bal^2}\bal\right)\cdot\bH.
\efr
It is a simple matter to verify \BOGE\ once one notices that
$[t_j,\xi^I]=0$. Furthermore the solution has the correct asymptotics
at spatial infinity:
$$
\lim_{\vert{\underline r}\vert
\rightarrow\infty}\Phi^I(\ur)=\lambda^I\left\vert\ba^I\cdot\bal\right
\vert{\bal\cdot\bH\over\bal^2}+\xi^I=\ba^I\cdot\bH.
\efr
The anti-monopole solution can be embedded in a similar way.

The magnetic charge of the solution is
$$
Q_M^I={\ba^I\cdot\bal\over\bal^2}=\ba^I\cdot\bal^\star,
\efr
where the co-root $\bal^\star=\bal/\bal^2$ is defind to be 
the magnetic charge vector of the solution.
The solution has a mass which saturates the Bogomol'nyi bound
$$
M={4\pi\over g}\left\vert\ba^I\cdot\bal^\star\right\vert.
\efr
Notice that the
solution can be rotated with an SO$(N_{\cal R})$
transformation. However this has the effect of rotating the Higgs VEV
$\ba^I$ so it actually relates solutions at {\it different\/} points
in the moduli space of vacua $W$.

For a theory with an $N=2$ supersymmetry, the R-symmetry is SO(2). In
this case, it will prove more convenient to rewrite the
Higgs field $\Phi^I$ as a single complex-valued field
$\Phi=\Phi^1+i\Phi^2$. In this
case the VEV on the sphere at spatial infinity is
$\ba\cdot\bH$, where now $\ba$ is a complex $r$-dimensional vector
and the Higgs field of the embedded monopole solution is
$$
\Phi(\ur)={\bal\cdot\ba\over\vert\ba\cdot\bal\vert}
\phi_j\left(\ur;\vert\ba\cdot\bal\vert\right)
t_j+\left(\ba-{\ba\cdot\bal\over\bal^2}\bal\right)\cdot\bH.
\nfr{EMC}

\chapter{The kinematics of monopole decay}

In this section we identify the kinematically allowed
regions in the classical moduli space of vacua $W$
where monopoles can decay. The treatment is a generalization of that
for the $N=2$ theory with SU$(n)$ gauge group appearing in [\Ref{FH1}].
Recall that a monopole associated to a 
root $\bal$ has a magnetic charge $\ba^I\cdot\bal^\star$. In
general because the state saturates the generalized Bogomol'nyi bound
it is below threshold for decay into a
multi-particle state of the same magnetic charge. However, on
submanifolds in the moduli space of vacua the monopole
can be at threshold for decay into other stable BPS states. In fact this
can occur whenever there exist two roots $\bgamma$ and $\bd$
such that
$$
\left\vert\ba^I\cdot\bal^\star\right\vert=
N\left\vert\ba^I\cdot\bgamma^\star\right\vert+
M\left\vert\ba^I\cdot\bd^\star\right\vert,
\nfr{CMS}
where conservation of magnetic charge requires 
$$
\bal^\star=N\bgamma^\star+M\bd^\star
\efr
and $N$ and $M$ are two positive integers $\geq1$. 
We denote the submanifold of $W$ on which \CMS\ is satisfied as the Curve
of Marginal Stability (CMS)
$C_{\bgamma,\bd}$. The condition \CMS\ is equivalent to
$$
\ba^I\cdot\bgamma=\lambda\ba^I\cdot\bd,\qquad\lambda>0,
\nfr{CMST}
or equivalently for a complex Higgs field
$$
{\ba\cdot\bgamma\over\ba\cdot\bd}\in{\Bbb R}>0.
\nfr{CMSF}
Notice that \CMS\ involves $N_{\cal R}-1$ real
condition, so the CMS are generically submanifolds of $W$ 
with co-dimension $N_{\cal R}-1$.

We now prove an important property of
$C_{\bgamma,\bd}$. The curve only has non-trivial overlap with the
interior of
$W$ if $\bgamma$ and $\bd$ are either both positive or both negative  
roots (with respect to the fundamental Weyl chamber defined by
$\ba^1$ or ${\rm Re}(\ba)$ in the case of a complex Higgs field). 
To see this consider the map $W\rightarrow S^1$ given by
$$
\cos\theta={(\bgamma\cdot\ba^I)(\ba^I\cdot\bd)\over
\left\vert\bgamma\cdot\ba^I\right\vert\left\vert\ba^I\cdot\bd\right\vert}.
\nfr{ECMS}
The CMS \CMST\ maps to
$\theta=0$. Suppose that $\bgamma$ is a positive root and $\bd$ is a
negative root, then since $\ba^1\cdot\bgamma>0$ and $\ba^1\cdot\bd<0$
the image of $W$ can never be $\theta=0$ which proves that
$C_{\bgamma,\bd}$ has no overlap with the interior of 
$W$ if $\bgamma$ is a positive
root and $\bd$ is a negative root, and vice-versa. On the contrary, if
$\bgamma$ and $\bd$ are both positive, or both negative, roots, then
the CMS \CMST\ can have non-trivial overlap with the interior of $W$.

So the following picture emerges. For each way of writing a positive co-root
$\bal^\star$ as the sum of two positive co-roots
$N\bgamma^\star+M\bd^\star$, for integers $N$ and $M$ both $\geq1$,
there exists a CMS on which the $\bal$
monopole is at threshold for decay into $N\times\bgamma$ plus
$M\times\bd$ monopoles. Of course, on the intersections of its
different CMS a
monopole can decay into states corresponding to more than two distinct
roots.

\chapter{Monopole zero modes for complex Higgs}

In this section we consider the structure of the moduli space of a
monopole ${\cal M}_{\rm mon}$. Our approach will be a generalization
of that of Weinberg [\Ref{WB1}] to the case of a complex Higgs field.
In particular, we shall determine under what conditions the zero modes
around the spherically symmetric solution discovered in [\Ref{WB1}] are
stabilized in the complex Higgs theory.

The dimension of the moduli space of a monopole is equal to the number
of normalizable zero modes around the monopole solution.
Rather than consider directly the equations for the zero modes
of the Higgs and gauge fields about a monopole solution, it is more
convenient, following Weinberg [\Ref{WB1}],
to consider the zero modes of the Dirac fermion field of the $N=2$ theory  
in the background of a monopole. The bosonic zero modes are then
related to the fermionic zero modes by the supersymmetry left unbroken
by the monopole background in such a way that for each
Dirac zero mode there are two bosonic, or Yang-Mills zero modes [\Ref{G1}].

The Dirac equation in the background of the monopole solution \EMC\ is
$$
\left[i\gamma^\mu D_\mu-{\rm Re}(\Phi)+i\gamma^5{\rm
Im}(\Phi)\right]\Psi=0.
\nfr{DIRAC}
The monopole solution in the gauge with $A_0=0$ is time-independent
and so we look for solutions of \DIRAC\ of the form
$\Psi(\ur,t)=e^{iEt}\Psi(\ur)$. Using a representation
$$
\gamma^j=\pmatrix{-i\sigma_j&0\cr0&i\sigma_j\cr},\qquad
\gamma_0=\pmatrix{0&-i\cr i&0\cr},\qquad\gamma_5=\pmatrix{0&1\cr1&0\cr},
\efr
we can write the equation for the modes as
$$
H\Psi=\pmatrix{{\rm Im}(\Phi)&i\sigma_jD_j+i{\rm Re}(\Phi)\cr
i\sigma_jD_j-i{\rm Re}(\Phi)&-{\rm Im}(\Phi)\cr}\Psi=E\Psi.
\nfr{ME}

The SO(2) R-symmetry acts on the fields in the following way:
$$
\Phi\mapsto e^{i\epsilon}\Phi,\qquad\Psi\mapsto
e^{i\epsilon\gamma^5/2}\Psi,\qquad A_i\mapsto A_i.
\nfr{RSYM}
Under such a
transformation the operator $H$ is {\it not\/} invariant because the
Higgs field of the background monopole solution
is not invariant. Such transformations, therefore, do not lead to a
degeneracy, however, we can use them to rotate the VEV
$\ba\rightarrow\ba'$ so that $\ba'\cdot\bal\in{\Bbb R}>0$. We will denote the
transformed real and imaginary part of the Higgs field as $P$ and $S$,
respectively. The transformation is achieved by choosing
the parameter $e^{i\epsilon}$ in \RSYM\ to
be $(\ba^*\cdot\bal)/\vert\ba\cdot\bal\vert$ and is explicitly
$$
\Phi\mapsto\Phi'=(P+iS)=
{\ba^*\cdot\bal\over\vert\ba\cdot\bal\vert}\Phi.
\nfr{HPS}
We will denote the transformed Higgs VEV as
$$
\ba'={\ba^*\cdot\bal\over\vert\ba\cdot\bal\vert}\ba.
\efr
The reason why this is useful is because
the transformed Dirac equation is now related in a simple way to 
that of a real Higgs theory since after the transformation, 
the real and imaginary parts of the Higgs field are  
$$\eqalign{
P&=\phi_j(\ur;\vert\ba\cdot\bal\vert)t_j+\left({\rm Re}(\ba')-
{\ba'\cdot\bal\over
\bal^2}\bal\right)\cdot\bH\cr
S&={\rm Im}(\ba')\cdot\bH.\cr}
\efr
So, bearing in mind that $\ba'\cdot\bal=\vert\ba\cdot\bal\vert$,
the real part of the Higgs field is identical to the Higgs field for a
monopole in a real Higgs theory with VEV ${\rm Re}(\ba')$ and the
imaginary part of the Higgs field is a constant. In the transformed
variables the Bogomol'nyi equations \BOGE, for time independent fields in
the $A_0=0$ gauge, become
$$
B_i=D_iP,\qquad D_iS=0,\qquad
[P,S]=0.
\nfr{SBO}

The transformed Dirac equation is 
$$
H'\Psi=\pmatrix{S&-{\cal D}^*\cr -{\cal D}&-S\cr}
\Psi=E\Psi,
\nfr{DIR}
where
$$
{\cal D}=-i\sigma_jD_j+iP,\qquad{\cal D}^*=
-i\sigma_jD_j-iP,
\efr
and $P$ and $S$ act by adjoint action on $\Psi$. We are now interested
in the zero energy solutions of \DIR. Fortunately, we
can draw on the results of Weinberg, who calculated the number of
zero modes of the two operators ${\cal D}$ and ${\cal
D}^*$. First of all, let us write $H'=H_0+K$, where
$$
H_0=\pmatrix{0&-{\cal D}^*\cr-{\cal D}&0\cr},\qquad K=\pmatrix{
S&0\cr 0&-S\cr}.
\efr
It follows from the Bogomol'nyi equations \SBO\ that
$$\eqalign{
{\cal D}^*{\cal D}&=-D_i^2+P^2-2i\sigma_iB_i\cr
{\cal D}{\cal D}^*&=-D_i^2+P^2.\cr}
\efr
So ${\cal D}{\cal D}^*$ is positive and has no non-trivial zero modes,
implying that ${\cal D}^*$ itself has no such modes. On the contrary
${\cal D}^*{\cal D}$ can have zero modes given by the zero modes of
${\cal D}$ itself. The number of such zero modes has been calculated
by Weinberg from the Callias index theorem [\Ref{WB1}]. We review the
results later in this section.

If $\psi$ is a zero mode of ${\cal D}$ then it follows that
$$
\left({\psi\atop0}\right),
\nfr{ZMD}
is a zero mode of $H_0$. Using $\{H_0,K\}=0$, it follows that
$$
H^{\prime2}=H_0^2+K^2=\pmatrix{{\cal D}^*{\cal D}+S^2&0\cr
0&{\cal D}{\cal D}^*+S^2\cr},
\efr
and since  $K^2$ is a positive operator the modes \ZMD\ are 
lifted if
$$
S\psi=[\left({\rm Im}(\ba')\cdot H\right),\psi]\neq 0.
\efr
To conclude,
the zero modes of $H'$, and hence $H$, 
are given by the subspace of the zero modes
of ${\cal D}$ with zero eigenvalue with respect to $S$.

In order to find the eigenvalues of the modes under $S$ we need a more
explicit description of the modes. Fortunately, the results we need
have been established by Weinberg [\Ref{WB1}]. First of all, the
real part of the VEV ${\rm Re}(\ba')$ defines a set of simple roots
$\bal'_i$ under which any positive root has a positive inner product
with ${\rm Re}(\ba')$. Notice that these simple roots are {\it not\/}
in general the same as $\bal_i$ defined in section 2 with respect to
$\ba^1\equiv{\rm Re}(\ba)$. To determine 
the number of zero modes of the monopole 
constructed via an $su(2)$ embedding \SUT, 
one expands the co-root $\bal^\star$ in terms of the simple co-roots
$\bal_i^{\prime\star}=\bal'_i/\bal_i^{\prime2}$:
$$
\bal^\star=\sum_{i=1}^rn_i\bal^{\prime\star}_i,
\efr
where the $n_i\in{\Bbb Z}\geq0$. The
overall number of Dirac zero modes is then 
$\left(2\sum_{i=1}^rn_i\right)$ and hence the number of bosonic, or 
Yang-Mills, zero modes is
$$ 
4\sum_{i=1}^rn_i.
\efr
The result has a simple interpretation: in general a monopole is a
composite object and the classical solution may be deformed into an
asymptotic region consisting of well separated fundamental monopoles
associated to the simple roots $\bal'_i$. 
For each fundamental monopole
there are four bosonic zero modes corresponding to the centre-of-mass
and overall U(1) charge rotation. 

Following Weinberg [\Ref{WB1}], we now expand the
adjoint valued field $\Psi$ in terms of the generators of $g$ in a
Cartan Weyl basis. The zero modes can be associated to multiplets of
generators under adjoint action by the $su(2)$ subalgebra of $g$ 
defined by the root $\bal$ in \SUT. The isospin
of a generator $E_\bb$ is given by 
$$
t_3={\bal\cdot\bb\over\bal^2},
\efr
and a multiplet is labelled by a total isospin $t$ along with 
$$
y={{\rm Re}(\ba')\cdot\bb\over{\rm Re}(\ba')\cdot\bal}-t_3,
\efr
which has the same value for any generator in the multiplet. The only
possibilities for the total isospin are $t=0,\half,1,\thalf$.
For each multiplet, 
the number of (normalizable) Dirac zero modes depends upon $t$
and $y$ in the following way [\Ref{WB1}]:
$$\eqalign{
t=\half:&\qquad 0<|y|<\half,\quad{\rm one}\cr
        &\qquad \half<|y|,\quad{\rm none}\cr
t=1:&\qquad 0\leq|y|<1,\quad{\rm two}\cr
        &\qquad 1<|y|,\quad{\rm none}\cr
t=\thalf:&\qquad 0<|y|<\half,\quad{\rm four}\cr
        &\qquad \half<|y|<\thalf,\quad{\rm three}\cr
        &\qquad \thalf<|y|,\quad{\rm none}.\cr}
\efr

We can physically identify each multiplet in the following
way. Firstly, there is always a $t=1$ multiplet with $y=0$ 
corresponding to the three generators of
the $su(2)$ defining the embedding \SUT. This gives rise to two Dirac,
i.e. four bosonic zeros modes, which reflect the three translational
and one U(1) charge rotational degrees-of-freedom of the spherically
symmetric $\bal$ monopole that are always present.

The other multiplets come in hermitian conjugate pairs with the
opposite value of $y$; hence, the
number of bosonic zero modes is always divisible by 4. Consider a
particular pair of conjugate multiplets. We can label this pair of
multiplets by $\bb$ the root whose generator
has $t_3=t$ in the multiplet with $y>0$. 
The number of bosonic zero modes associated to the pair of
multiplets is equal to $4K$ for some integer $K\geq0$. We will verify below
on a case-by-case basis for each pair of multiplets $(t,\pm y)$
that the co-root $\bal^\star$ can always be written
$$
\bal^\star=N\bgamma^\star+M\bd^\star,
\nfr{DECA}
where $\bgamma$ and $\bd$ are two positive roots with respect to $\bal'_i$
and $N$ and $M$ are two 
positive integers with $N+M=K+1$. The interpretation of these $4K$ zero modes
is now apparent, they correspond to the freedom to deform the
spherically symmetric $\bal$ monopole into an asymptotic region which
consists of $N\times\bgamma$ plus $M\times\bd$ monopoles. Such a
configuration requires $4(N+M-1)=4K$ relative 
degrees-of-freedom corresponding
to relative positions and U(1) charge angles 
of the $N+M$ consistent monopoles. These
degrees-of-freedom are manifested in the $4K$ zero modes coming from
the pair of multiplets associated to $\bb$.

We now establish \DECA\ on a case-by-case basis. In the following,
$\bal$ is always the root defining the $su(2)$ embedding and $\bb$ is
the root labelling the generator with $t_3=t$ and $y>0$ of a pair of 
$(t,\pm y)$ multiplets. We need not consider the $(t=1,y=0)$ multiplet
associated with the centre-of-mass and overall U(1) charge rotation
that is always present.

(i) $t=\half$, which requires $\bal$ to be a long root. 
In this case the pair of multiplets consists of generators
$$
\left\{E_\bb,E_{\bb-\bal}\right\}\qquad{\rm and}\qquad
\left\{E_{\bal-\bb},E_{-\bb}\right\}.
\efr
There are two separate cases to consider depending on whether $\bb$ is
a long or a short root. 

If $\bb$ is a long root (which covers all the
simply-laced cases) then
$$
\bal=\bb+(\bal-\bb)\qquad\Rightarrow\qquad\bal^\star=
\bb^\star+(\bal-\bb)^\star.
\nfr{WFF}
On the other hand, 
if $\bb$ is a short root, which can only occur in a non-simply-laced
algebra, then
$$
\bal=2(\bal-\bb)+(2\bb-\bal)\qquad\Rightarrow\qquad\bal^\star=
(\bal-\bb)^\star+(2\bb-\bal)^\star.
\efr

In both cases, the condition that $0<\vert y\vert<\half$ is equivalent to the
requirement that
$$
{\rm Re}(\ba')\cdot(\bal-\bb)>0\qquad{\rm and}\qquad
{\rm Re}(\ba')\cdot(2\bb-\bal)>0,
\efr
and therefore implies that $\bb$, $\bal-\bb$ and $2\bb-\bal$ 
are all positive
roots with respect to ${\rm Re}(\ba')$, as required.
The number of bosonic zero modes in both cases is 4 which is consistent with 
a decay into 2 stable monopoles.

(ii) $t=1$, which requires $\bal$ to be a short root. 
In this case the pair of multiplets consists of generators
$$
\left\{E_\bb,E_{\bb-\bal},E_{\bb-2\bal}\right\}\qquad{\rm and}\qquad
\left\{E_{2\bal-\bb},E_{\bal-\bb},E_{-\bb}\right\}.
\efr
In this case $\bb$ is a long root and
$$
\bal=(2\bal-\bb)+(\bb-\bal)\qquad\Rightarrow\qquad\bal^\star=
2(2\bal-\bb)^\star+(\bb-\bal)^\star.
\nfr{WSS}
The condition that $0<\vert y\vert<1$ is equivalent to the
requirement that
$$
{\rm Re}(\ba')\cdot(2\bal-\bb)>0\qquad{\rm and}\qquad
{\rm Re}(\ba')\cdot(\bb-\bal)>0,
\efr
and therefore implies that both $2\bal-\bb$ and $\bb-\bal$ are positive
roots with respect to ${\rm Re}(\ba')$, as required.
The number of bosonic zero modes is 8 which is consistent with 
a decay into 3 stable monopoles.

(iii) $t=\thalf$, which requires $\bal$ to be a short root. (This
example only occurs in the case when the gauge group is $G_2$.)
In this case the pair of multiplets consists of generators
$$
\left\{E_\bb,E_{\bb-\bal},E_{\bb-2\bal},E_{\bb-3\bal}
\right\}\qquad{\rm and}\qquad
\left\{E_{3\bal-\bb},E_{2\bal-\bb},E_{\bal-\bb},E_{-\bb}\right\}.
\efr
In this case $\bb$ is a long root and there are two possible decays
$$\eqalign{
\bal=(2\bb-3\bal)+2(2\bal-\bb)&\qquad\Rightarrow\qquad\bal^\star=
3(2\bb-3\bal)^\star+2(2\bal-\bb)^\star\cr
\bal=(3\bal-\bb)+(\bb-2\bal)&\qquad\Rightarrow\qquad\bal^\star=
3(3\bal-\bb)^\star+(\bb-2\bal)^\star.\cr}
\nfr{WGT}
It is important in the above that all the terms in brackets are
actually roots of the algebra.
The first expression for $\bal^\star$ in \WGT\  is relevant to the case when
$0<\vert y\vert<\half$ which is equivalent to the 
requirement that
$$
{\rm Re}(\ba')\cdot(2\bb-3\bal)>0\qquad{\rm and}\qquad
{\rm Re}(\ba')\cdot(2\bal-\bb)>0,
\efr
and therefore implies that both $2\bb-3\bal$ and $2\bal-\bb$ are positive
roots with respect to ${\rm Re}(\ba')$, as required.
The number of bosonic zero modes is 16 which is consistent with 
a decay into 5 stable monopoles.

The second expression in for $\bal^\star$ \WGT\  is relevant to the case when
$\half<\vert y\vert<\thalf$ which is equivalent to the 
requirement that
$$
{\rm Re}(\ba')\cdot(\bb-2\bal)>0\qquad{\rm and}\qquad
{\rm Re}(\ba')\cdot(3\bal-\bb)>0,
\efr
and therefore implies that both $\bb-2\bal$ and $3\bal-\bb$ are positive
roots with respect to ${\rm Re}(\ba')$, as required.
The number of bosonic zero modes is 12 which is consistent with 
a decay into 4 stable monopoles.

Although the explicit expression for $\bgamma$ and $\bd$ can only be
written down on a case-by-case basis, it is important that in all
cases, as
one can find by inspecting \WFF, \WSS\ and \WGT, 
$\bgamma$ and $\bd$ can always be expressed as a linear combination 
of the two vectors $\bal$ and $\bb$. This fact
will play a crucial role in our argument below. Another important
consistency check of the picture is that the number of zero modes
always matches the degrees-of-freedom implied by the number of decay
products. 
 
Now that we have described the zero modes of the real Higgs model, the
next question concerns their fate when a complex Higgs theory is
considered. Recall, that a zero mode is lifted if the
eigenvalue of ${\rm Im}(\Phi)={\rm Im}(\ba')\cdot\bH$ on the mode is
non-zero. The eigenvalue associated to the $\bb$ multiplet is
$$
{\rm Im}(\ba')\cdot\bb={1\over\vert\ba\cdot\bal\vert}{\rm
Im}\left((\ba^*\cdot\bal)(\bal\cdot\bb)\right),
\efr
and the conjugate multiplet has the opposite eigenvalue.

Notice that the $t=1$ and $y=0$ multiplet associated to the embedded
$su(2)$ \SUT\ is never lifted because ${\rm Im}(\ba')\cdot\bal=0$. This was
to be expected since is corresponds to the freedom to shift the
centre-of-mass and perform an overall 
U(1) charge rotation on the solution. The
other multiplets are generally lifted and so the
spherically symmetric monopole solution generically has no additional zero
modes. However, on curves of co-dimension one in the moduli space of
vacua, whenever ${\rm Im}(\ba')\cdot\bb=0$, for some root $\bb$ with
$\bb\cdot\bal\neq0$, then
additional zero modes appear that describe the freedom to deform the
solution into $K+1$ well separated constituent stable 
monopoles. On this special
curve the $\bal$ monopole is at threshold for the decay:
$$
\left\vert\bal^\star\cdot\ba\right\vert=
N\left\vert\bgamma^\star\cdot\ba\right\vert
+M\left\vert\bd^\star\cdot\ba\right\vert,
\efr
or
$$
{\bgamma\cdot\ba'\over\bd\cdot\ba'}\equiv
{\bgamma\cdot\ba\over\bd\cdot\ba}
\in{\Bbb R}\geq0.
\nfr{CMAG}
This follows from the fact that
since $\bgamma$ and $\bd$ can be expanded in terms of $\bal$ and $\bb$ 
and ${\rm Im}(\ba')\cdot\bal={\rm Im}(\ba')\cdot\bb=0$ necessarily implies
${\rm Im}(\ba')\cdot\bgamma={\rm Im}(\ba')\cdot\bd=0$.
Notice that \CMAG\ coincides with the definition of $C_{\bgamma,\bd}$ 
in \CMSF.

There is one remaining point to clear up. The decay products
$\bal^\star=N\bgamma^\star+M\bd^\star$ that are associated to the two 
roots $\bgamma$
and $\bd$, are required to be positive roots with respect to the simple roots
$\bal'_i$ defined with respect to ${\rm Re}(\ba')$. 
However in section 3 we
determined that the decay can only proceed if the decay products where
associated to positive roots with respect to the simple roots $\bal_i$
defined with respect to $\ba^1\equiv{\rm Re}(\ba)$. This is consistent
because precisely on the CMS the two
roots $\bgamma$ and $\bd$ are positive roots with respect to {\it both\/}
definitions of simple root. To see this consider
$$
{\rm Re}(\ba')\cdot\bgamma={1\over\vert\ba\cdot\bal\vert}
{\rm Re}\left((\ba\cdot\bgamma)(\ba^{*}\cdot\bal)\right)
={\bal^2\over\vert\ba\cdot\bal\vert}{\rm
Re}\left(N{\vert\ba\cdot\bgamma\vert^2\over\bgamma^2}+M{(\ba\cdot
\bgamma)(\ba^{*}
\cdot\bd)\over\bd^2}\right),
\efr
But on the CMS curve $(\ba\cdot\bgamma)(\ba^{*}
\cdot\bd)$ is a positive real number, therefore we deduce that ${\rm
Re}(\ba')\cdot\bgamma$ is a positive number and consequently on the
$C_{\bgamma,\bd}$ the root $\bgamma$ is also a positive
root with respect to the simple roots $\bal'_i$. An identical argument
follows for $\bd$.

\chapter{The dynamics of monopole decay}
  
In section 3, we considered the CMS of the classical moduli
space of vacua $W$ where it was energetically possible for dyons to
decay. In the present section we now pose the dynamical question as to whether
dyons do actually decay on these CMS. Although some pieces of the 
argument have not been fully proven for all possible gauge groups
the resulting picture is rather convincing: in an $N=2$ pure
gauge theory at weak coupling dyons
always decay on their CMS, whereas in the related 
$N=4$ theory they do not
decay. This is one manifestation of the relative
simplicity of $N=4$ theories over $N=2$ theories.

The analysis in the last section has all been at the level of zero
modes around the spherically symmetric monopoles solutions. 
From this, we have concluded that these solutions, 
obtained by embeddings of the SU(2) monopole, are
generically stable in $W$. The only zero modes which are generically
present account for the centre-of-mass and U(1) charge angle of the
solution. On the submanifold \CMAG, of co-dimension one, $4(N+M-1)$
modes are destabilized and become zero modes. These modes reflect that
fact that the monopole moduli space includes deformations away from
the spherically symmetric solution leading to an enlargement of the 
moduli space by a factor ${\cal M}_0$ of dimension $4(N+M-1)$.
In the case when $N=M=1$ corresponding to zero modes with $t=\half$,
${\cal M}_0$ is known explicitly to be a Euclidean
Taub-NUT manifold [\Ref{LWY2},\Ref{GL},\Ref{LWY1}]. However for $t=1$
or $t=\thalf$, the dimension of ${\cal M}_0$ is 8 and 12 or 16,
respectively, and the manifold is only known in asymptotic regions
[\Ref{LWY2}]. In all cases, however, ${\cal M}_0$ is a hyper-K\"ahler
manifold [\Ref{G1}].

We now make the hypothesis that the spectra of $N=4$ theories manifest
exact GNO duality for all gauge groups. 
The idea is that at all points
in the moduli space of vacua in an $N=4$ supersymmetric gauge theory, 
the spectrum
of monopoles must match the spectrum of gauge bosons in the dual
theory, defined as an $N=4$ supersymmetric gauge theory but whose
gauge group has a Lie algebra $g^\star$
being the ``dual'' of the original Lie algebra $g$. 
The dual algebra is defined as the algebra whose simple
roots are $\bal_i^\star=\bal_i/\bal_i^2$, up to some overall
normalization. Hence,
all the simply-laced theories are self-dual, whereas
the algebra/dual algebra relationships of the non-simply-laced theories are
$$\eqalign{
sp(n)&\leftrightarrow so(2n+1)\cr
f_4&\leftrightarrow f_4'\cr
g_2&\leftrightarrow g_2',\cr}
\efr
where the prime indicates a re-labelling of the roots.
Since in the dual theory there is a
gauge boson associated to every root $\bal^\star$ of $g^\star$, with electric
charge $Q_E^I=\ba^I\cdot\bal^\star$, for GNO
duality to be reflected in the spectrum
there must be a monopole of magnetic charge $Q_M^I=\ba^I\cdot\bal^\star$
at all points in the moduli space of vacua $W$.
In other words, in
the $N=4$ theory monopoles do not decay on their CMS and consequently
they must exist on these submanifolds
as bound-states at threshold. Such states can indeed
exists within the semi-classical approximation if there exists a
square-integrable
harmonic form on each ${\cal M}_0$. In the cases where 
${\cal M}_0$ is known (for $N=M=1$ i.e. $t=\half$)
the appropriate harmonic form has been found
[\Ref{LWY2},\Ref{GL},\Ref{LWY1},\Ref{GIB1}], thus
providing strong evidence for GNO duality.

If GNO duality is exact in the associated $N=4$ theory, i.e. the
bound-states at threshold exist, it
implies that the corresponding dyons in an $N=2$ theory
actually decay on their CMS. 
To appreciate this, we have to consider the difference between the
quantum mechanics that arises on ${\cal M}_0$ in the context of the
semi-classical approximation of monopoles in an $N=2$
and $N=4$ supersymmetric gauge theory, respectively. 
In both cases the space-time supersymmetry is
manifested by supersymmetries in the quantum mechanics on ${\cal
M}_0$. For $N=4$ there are ``$4\times1/2$'' supersymmetries [\Ref{B1}]
whilst for $N=2$
there are ``$2\times1/2$'' supersymmetries [\Ref{G1}]. 
In both cases states can be
associated to differential forms on ${\cal M}_0$, however for $N=4$
the states correspond to any form whilst in $N=2$ they correspond only
to holomorphic forms.\note{Recall that ${\cal M}_0$ is a (hyper) K\"ahler
manifold.} Bound-states correspond to normalizable 
ground-states of the quantum mechanics
which in turn correspond to square-integrable harmonic forms on ${\cal
M}_0$. As we have argued, exact duality implies that there is a unique
square-integrable harmonic form on ${\cal M}_0$ and 
since a holomorphic harmonic form would inevitably have an
anti-holomorphic partner, we conclude that there cannot be 
a square-integrable holomorphic harmonic form on ${\cal M}_0$.
So there are no bound-states of monopoles in an $N=2$ theory (with no
matter) and the
dyons in these theories must decay on their CMS at weak coupling. 
As we have mentioned,
the statement that monopoles in the $N=2$ theory actually decay can
really only be proved for decays with $N=M=1$, i.e. when there are
only two dyons in the final state. Nevertheless, this is enough to prove that
monopoles decay in {\it all\/} the simply-laced theories, where,
generically, all the decays have two dyons in the final state.

Exactly the same line of argument should prove that there are no stable
dyons
in an $N=2$ theory whose magnetic charge $Q_M^I=\ba^I\cdot\bb^\star$
are given by vectors $\bb$ which are not roots of $g$. Such states
do exist in the associated $N=4$ theory, with $\bb=n\bal$, where
$\bal$ is a root of $g$ and $n\in{\Bbb Z}>1$ [\Ref{GL},\Ref{DFHK2}]. 
Given the uniqueness of
these bound-states in the $N=4$ theory, which follows from the
analysis in the SU(2) theory [\Ref{SEN},\Ref{POR}], they will not occur
in the $N=2$ theory.

\bjump
I would like to thank PPARC for an Advanced Fellowship.

\references

\beginref
\Rref{SW1}{N. Seiberg and E. Witten, Nucl. Phys. {\bf B426} (1994) 19,
{\tt hep-th/9407087}}
\Rref{SW2}{N. Seiberg and E. Witten, Nucl. Phys. {\bf B431} (1994) 484,
{\tt hep-th/94080997}}
\Rref{BF}{A. Bilal and F. Ferrari, Nucl. Phys. {\bf B469} (1996) 387,
{\tt hep-th/9602082}} 
\Rref{F1}{A. Fayyazuddin, {\tt hep-th/9504120}}
\Rref{AFS}{P.C. Argyres, A.E. Faraggi and A.D. Shapere, {\tt hep-th/9505190}}
\Rref{GNO}{P. Goddard, J. Nuyts and D. Olive, Nucl. Phys. {\bf B125}
(1977) 1}
\Rref{SEN}{A. Sen, Phys. Lett. {\bf B329} (1994) 217}
\Rref{POR}{M. Porrati, Phys. Lett. {\bf B377} (1996) 67, {\tt
hep-th/9505187}
\newline G. Segal and A. Selby, Commun. Math. Phys. {\bf177} (1996) 775}
\Rref{FH1}{C. Fraser and T.J. Hollowood, {\tt hep-th/9610142}}
\Rref{MAN}{N. Manton, Phys. Lett. {\bf B110} (1982) 54}
\Rref{TP}{G. 't Hooft, Nucl. Phys. {\bf B79} (1976) 276\newline
A.M. Polyakov, JETP Lett. {\bf20} (1974) 194}
\Rref{WB1}{E.J. Weinberg, Nucl. Phys. {\bf B167} (1980) 500;  
Nucl. Phys. {\bf B203} (1982) 445}
\Rref{GL}{J.P. Gauntlett and D. A. Lowe, Nucl. Phys. {\bf B472} (1996) 194,
{\tt hep-th/9601085}}
\Rref{LWY3}{K. Lee, E.J. Weinberg and P. Yi, Phys. Rev. {\bf D54}
(1996) 6351, {\tt hep-th/9605229}}
\Rref{LWY1}{K. Lee, E.J. Weinberg and P. Yi, Phys. Lett. {\bf B386}
(1996) 97, {\tt hep-th/9601097}}
\Rref{LWY2}{K. Lee, E.J. Weinberg and P. Yi, 
Phys. Rev. {\bf D54} (1996) 1633, {\tt hep-th/9602167}}
\Rref{MUR}{M. Murray, {\tt hep-th/9605054}}
\Rref{GC}{G. Chalmers, {\tt hep-th/9605182}}
\Rref{GIB1}{G. Gibbons, Phys. 
Lett. {\bf B382} (1996) 53, {\tt hep-th/9603176}}
\Rref{GIB2}{G.W. Gibbons and P. Rychenkova, {\tt hep-th/9608085}}
\Rref{SC}{S.A. Connell, `The Dynamics of the ${\rm SU}(3)$ $(1,1)$
Magnetic Monopole', Unpublished preprint available by anonymous
ftp from : {\tt <ftp://maths.adelaide.edu.au/pure/-
mmurray/oneone.tex>}}
\Rref{G1}{J.P. Gauntlett, Nucl. Phys. {\bf B411} (1994) 443}
\Rref{GNO}{P. Goddard, J. Nuyts and D. Olive, Nucl. Phys. {\bf B125}
(1977) 1}
\Rref{PS}{M.K. Prasad and C.M. Sommerfield, Phys. Rev. Lett. {\bf35}
(1975) 760}
\Rref{DFHK1}{N. Dorey, C.Fraser, T.J. Hollowood and M.A.C. Kneipp, 
{\tt hep-th/9512116}}
\Rref{DFHK2}{N. Dorey, C. Fraser, T.J. Hollowood and M.A.C. Kneipp,
Phys. Lett. {\bf B383} (1996) 422, {\tt hep-th/9605069}}
\Rref{BAIS}{F.A. Bais, Phys. Rev. {\bf D18} (1978) 1206}
\Rref{B1}{J.D. Blum, Phys. Lett. {\bf B333} (1994) 92, 
{\tt hep-th/9401133}}
\Rref{H1}{M. Henningson, Nucl. Phys. {\bf B461} (1996) 101}
\Rref{AY}{O. Aharony and S. Yankielowicz, Nucl.Phys. {\bf B473} (1996)
93, {\tt hep-th/9601011}} 
\endref

\ciao